\documentclass{aastex}
\usepackage{spr-astr-addons}
\usepackage{url}

\RequirePackage{color}

\usepackage{bm}
\usepackage{amsmath}
\usepackage{amsfonts}
\usepackage{amssymb}
\usepackage{multirow}
\usepackage[misc]{ifsym}
\usepackage{cases}

\begin{document}

\title{Two Dimensional Classification of the \textit{Swift}/BAT GRBs}

\author{E. B. Yang\altaffilmark{1}}
\and
\author{Z. B. Zhang\altaffilmark{1,2,{*}}}
\and
\author{X. X. Jiang\altaffilmark{1}}

\altaffiltext{1}{Guizhou University, Department of Physics, College of Sciences,
Guiyang 550025, China}
\altaffiltext{2}{Key Laboratory for the Structure and Evolution of Celestial Objects,
	Chinese Academy of Sciences, Kunming, 650011, China}
\altaffiltext{*}{ E-mail: sci.zbzhang@gzu.edu.cn}

\begin{abstract}
Using Gaussian Mixture Model and Expectation Maximization algorithm, we
have performed a density estimation in the framework of $T_{90}$  versus
hardness ratio for 296 Swift/BAT GRBs with known redshift. Here,  Bayesian
Information Criterion has been taken to compare different models. Our
investigations show that two instead of three or more Gaussian components
are favoured in both the observer and rest frames. Our key findings
are consistent with some previous results.

\end{abstract}

\keywords{gamma-ray burst:general-- methods: data analysis --methods: statistical}

\section{Introduction}

The mystery
of Gamma Ray Bursts (GRBs) classification has puzzled astronomers
since 1990s. \cite{kouveliotou} firstly discovered the bimodal distribution
of $\log T_{90}$ and proposed the duration classification criterion, that is
long GRBs with $T_{90}>2s$ and short GRBs with $T_{90}<2s$, This classification
method has been widely accepted by many authors so far. The two kinds of GRBs
are thought to have different physical origins. Short GRBs are believed to be
the product of merging of binary systems, like two neutron stars (NS-NS) or a
system of neutron star and black hole (NS-BH)
\citep{nakar,zhangbing2009,gehrels09}.
LGRBs have close relationship with the collapse of massive stars that has been
mostly confirmed by observations \citep{woosley}.
However, \cite{horvath1998} analysed the $\log{T_{90}}$ distribution
of BATSE 3B catalogue including 797 GRBs and found
a possible existence of one more class. The phenomenal evidence was also found
in the datasets of \textit{BeppoSAX} \citep{horvath2009},
\textit{Swift}/BAT \citep{horvath2008}, \textit{Fermi}/GBM \citep{tarno2015aa} etc.

Besides the observed duration distribution, several authors also did similar
studies of the distribution in the rest frame. \cite{zhangzb2008} analysed the
duration distribution of 95 redshift-known \textit{Swift} GRBs and found that
the intrinsic duration is still bimodal. \cite{zitouni} found that 3
groups were statistically better that 2 groups after studying the intrinsic
duration distribution of a larger datasets redshift-known \textit{Swift} GRBs,
which contains 248 GRBs. \cite{tarno2016na} analysed the distribution
of 947 \textit{Swift} GRBs' observed duration and 347 redshift-known GRBs'
duration in both observer and rest frames. It is found that 3 groups are
statistically better than 2 groups in the observer frame (see also \citealt{tarno2016mn}).
In the rest frame, 2 groups are good enough for the 347 redshift-known GRBs.
Recently, \cite{horvath2016} studied the duration distribution 
of 888 \textit{Swift} GRBs observed before October 2015. They found that three 
log-normal function is better to fit the duration distribution 
than two log-normal function,
with a 99.9999\% significance level. Their results showed that the 
relative frequencies were 8\%, 35\% and 57\% for short, intermediate and long bursts, 
respectively (see also \citealt{horvath2008}). They further claimed that no 
significant differences of the redshift distribution were found between the 
intermediate GRBs and the long ones. However, the existence of this third class 
is still controversial so far.

Considering this, durations-based classification of GRBs might not be an only
standard. \cite{zhangbing2011} proposed that more observational properties,
namely \textit{redshift, fluence, flux, $E_{peak}$} of the $\upsilon f_{\upsilon}$
spectrum and etc., should be taken into account together. Meanwhile, SGRBs
usually have harder spectra than LGRBs \citep{kouveliotou,zhangzb2008}, the
clustering phenomenon in the plane of duration versus harness ratio
 ($T_{90}$ vs $HR$) should also be useful to classify the GRBs.
\cite{horvath2006,horvath2010} did  clustering analysis of $T_{90}$ vs $HR$
diagram for BATSE and \textit{Swift} datasets, respectively. They found the
possible existence of the third class. It is worth noting that the above
studies are carried out based on the observations only that may involve the
selection effect or bias. In addition, different methods applied for the same
sample may reach controversial judgement of GRB classes. To avoid all the above
influences, we apply Gaussian Mixture Model (GMM) and Expectation
Maximization (EM) Algorithm to the density estimation of $T_{90}$ vs $HR$
diagram for the redshift-known \textit{Swift} GRBs  in both the observer frame and
the rest frame. Note that the sample in this work only comprise of those 
\textit{Swift}/BAT GRBs with measured redshift, unlike some previous studies. 
This paper is organized as follows, Section~\ref{sect2} will give the description of our datasets and analysing
method. Our main results will be displayed in Section~\ref{sect3}.

\section{Data Preparation and Method}\label{sect2}

\subsection{Data Preparation}
\begin{figure*}[!ht]
	\includegraphics[width=2\columnwidth]{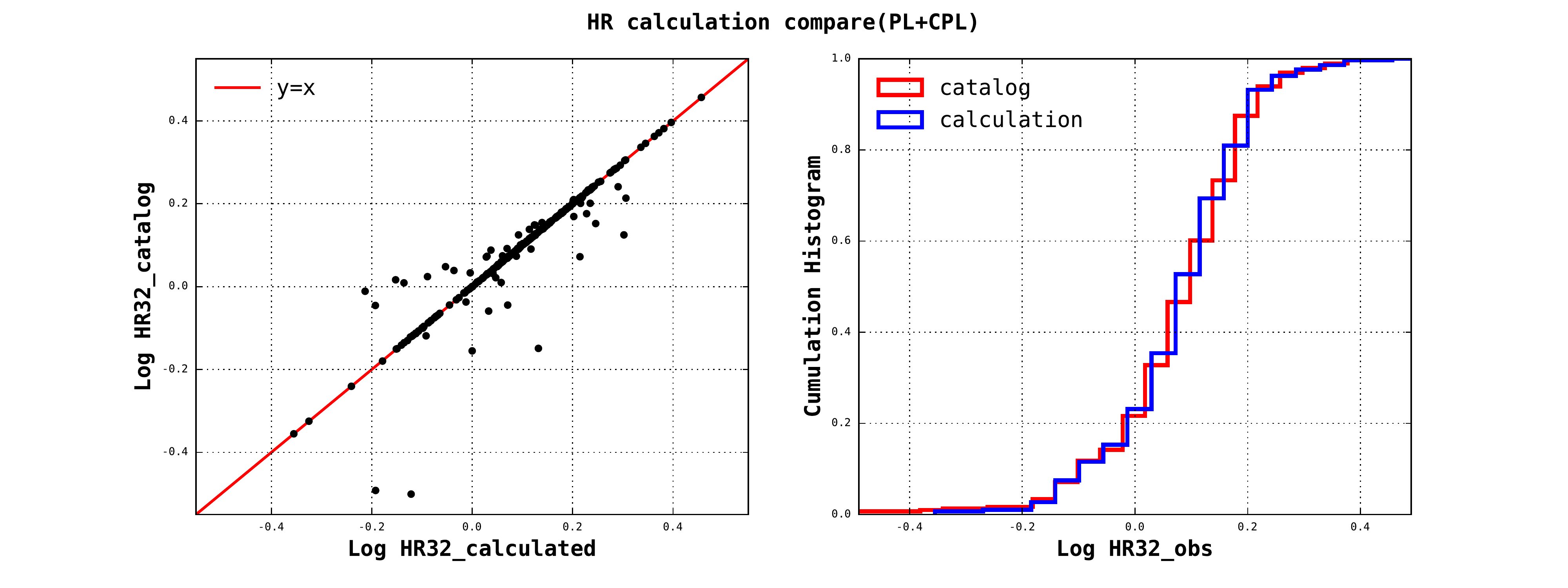}
	\caption{Comparing $HR_{32,obs}$ between our calculation(x axis)
	and the value(y axis) listed in the catalog on the left panel. The red
	solid line is the equality line. Their cumulation histograms are compared
	on the right panel.}\label{hr32_comp}
\end{figure*}
\begin{figure}[!ht]
	\includegraphics[width=1\columnwidth]{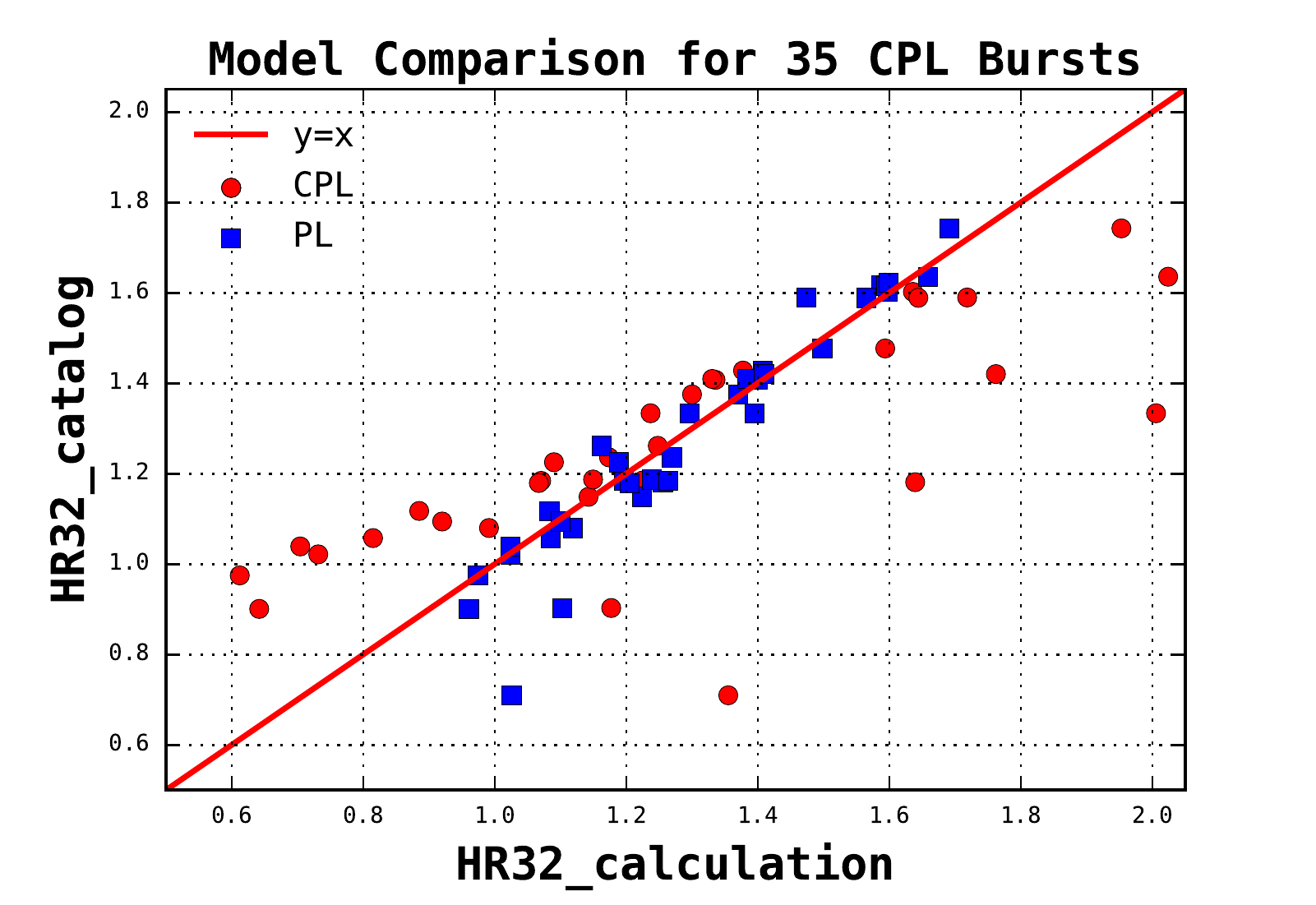}
	\caption{Comparing $HR_{32,obs}$ between the calculated values
	under PL model and CPL model with the values in catalog for the 35 CPL
	bursts. Blue squares are calculated by using the PL parameters and red
	solid points are calculated by using the CPL parameters}\label{cpl_comp}
\end{figure}
The current catalog of \textit{Swift}\footnote{\url{http://heasarc.gsfc.nasa.gov/W3Browse/swift/swiftgrb.html}}
 \citep{donato2012} contains all GRBs observed by \textit{Swift} from the
beginning of the mission, 20 Nov 2004 to 31 Dec 2012, of which 296 bursts have
both $T_{90}$, redshift, spectral properties (photon index $\alpha$,
\textit{fluence} and etc.). The observed spectral hardness $HR_{32,obs}$ in our
analysis was defined as the fluence ratio between 50 keV$\sim$100 keV and 25
keV$\sim$50 keV and it can be calculated by the fluence properties listed in
the catalog. To calculate the intrinsic spectral hardness
$HR_{32,int}$, we use the best fitted spectral model provided in the
\textit{Swift} official site. In our datasets, there are 261 GRBs following the
single power law model (PL) and other 35 GRBs are better to be described with
the cutoff power law model (CPL). The PL and CPL forms can be written as

\begin{numcases}
{f(E)=}
A*E^{-\alpha}\\
A*E^{-\alpha}*e^{-E/\beta}
\end{numcases}
where $\alpha$ are the single power law or the cutoff power law photon
index, $\beta$ is the peak energy, and $f(E)$ represents the photon flux at a
given energy. The  hardness ratio can be calculated by
\begin{equation}\label{hr_obs}
	HR_{32,obs}=\frac{\intop_{50keV}^{100keV}f(E)
		\cdot EdE}{\intop_{25keV}^{50keV} f(E) \cdot EdE}
\end{equation}
\begin{equation}\label{hr_res}
HR_{32,int}=\frac{\intop_{50*(1+z)keV}^{100*(1+z)keV}f(E')
\cdot E'dE'}{\intop_{25*(1+z)keV}^{50*(1+z)keV} f(E') \cdot E'dE'}
\end{equation}
where $z$ is the cosmological redshift. It is well known in the
literature that, with single power law spectrum, the hardness ratio remains the
same at any redshift. In the real situation, the hardness ratio should
depending on redshift\citep{bagoly}. The spectrum in the rest frame may has got
a more complicated form. Considering this, our calculation of hardness ratio in
the rest frame, especially for the 261 PL bursts, is simply an approximation of
the real value. We also calculate the ${HR}_{32,obs}$ by always using the best
fitted spectral model and compare it with that in the catalog in
Fig.~\ref{hr32_comp}. One can find that some points are obviously deviate from
the red solid equality line. Considering this, we check the 35 GRBs that follow
CPL model and compare the calculated $HR_{32,obs}$ under both PL model and CPL
model with the values in catalog. We compare the two models  in
Fig.~\ref{cpl_comp}, in which the PL form seems more consistent with
observations than the CPL one. In order to diagnose if the different model can
change the classification based on the $T_{90}$ distributions, we will do a
comparative study with the following method.

\subsection{Analysis Method}

We used the GMM and EM algorithm to estimate the density  of \textit{Swift} bursts with
known redshift in the $T_{90}-HR_{32}$ plane. The GMM
is a parametric probabilistic function representing a weighted sum
of Gaussian components. For the two-dimensional case,
the likelihood will be

\begin{equation}
P(\bm{{X}}|\omega,\bm{{\mu}},\bm{{\Sigma}})=\sum_{j=1}^{N}({\sum_{i=1}^{k}{\omega_{i}{N(\bm{{x_{j}}}|\bm{{\mu_{i}}},\bm{{\Sigma_{i}}})}}})
\end{equation}
where $\bm{{X}}=\{\bm{{x_{1},x_{2},...,x_{j}}}\}$ is the dataset,
$\omega_{i}$ is weight of the $i_{th}$ Gaussian component, $\bm{{\mu_{i}}}$
and $\bm{{\Sigma_{i}}}$ are the mean vector and the covariance
matrix correspondingly. $N(\bm{{x_{j}}}|\bm{{\mu_{i}}},\bm{{\Sigma_{i}}})$
is the density of the $i_{th}$ Gaussian component that can be calculated
by
\begin{multline}
N(\bm{{x_{j}}}|\bm{{\mu_{i}}},\bm{{\Sigma_{i}}})=
\frac{1}{2\pi}\\\cdot\frac{1}{\sqrt{\bm{\Sigma_{i}}}}\cdot\exp\left\{-\frac{1}{2}(\bm{{x_{j}-\mu_{i}}^{T}})\bm{\Sigma_{j}^{-1}}(\bm{x_{j}-\mu_{i}})\right\}
\end{multline}
The best parameters of an input model can be gained by using the EM algorithm,
which will maximize $\ln P(\bm{X}|\omega,\bm{\mu},\bm{\Sigma})$.
In our analysis, $\bm{\mu}$ contains two numbers
with the order of ($\log T_{90}, \log HR_{32}$). The GMM will assign a sample to
the Gaussian component with the highest probability.

Bayesian Information Criterion (BIC) \citep{bic,liddle2007} were usually used to determine the number of Gaussian
components. The BIC value is defined as

\begin{equation}\label{eq_bic}
BIC=p\ln N-2\ln P_{max}
\end{equation}
where $p$ is the number of parameters, $P_{max}$ is
the maximum likelihood. The BIC will choose the model when Eq.~\ref{eq_bic} reaches the
minimum $BIC_{min}$.

\begin{equation}
\Delta=BIC-BIC_{min}
\end{equation}

For other candidate models, if $0<\Delta<2$, they are also supported. If
$2<\Delta<6$, they are weakly supported. But if $\Delta>6$, these models are
 will not be reliable any more \citep{burnham2004}.

All our analyses are performed under \textit{scikit-learn}\footnote{\url{http://scikit-learn.org}}
\citep{sklearn}, a Python Machine Learning package. The histogram figures were
plotted following Knuth bin rule \citep{knuth} with the implement in
\textit{astropy}\footnote{\url{http://www.astropy.org}}.
\begin{table}[!ht]\centering
	\renewcommand{\arraystretch}{1.5}
	\caption{Optimized parameters for redshift-known {\it Swift} samples in both observer and rest frames. All values are calculated by using mixed spectrum model, PL or CPL. $\bm{\mu}$ contains two number with the order of ($\log T_{90}, \log HR_{32}$).}
	\scalebox{0.55}{
		\begin{tabular}{lcccccc}\hline
			&&&\textbf{PL+CPL}&&&\\ \hline
			Frame &  $\omega$   &$\bm{\mu}$          & $\bm{\Sigma}$   &BIC  &${\Delta}$	   &Model     \\
			\hline
			\multirow{12}{*}{\bf Observer}& 0.427              &(1.121, 0.043)
			&  $\left(\begin{array}{cc}
			0.583 &-0.054\\
			-0.054  & 0.023
			\end{array}\right)$
			&\multirow{4}{*}{222.755}&\multirow{4}{*}{0} &\multirow{4}{*}{2-G}        \\
			&0.573				&(1.953, 0.140)
			& $\left(\begin{array}{cc}
			0.162 &-0.007\\
			-0.007  & 0.006
			\end{array}\right)$
			&& \\
			&&&&&&\\
			& 0.026           &(-0.542,0.353)
			&$\left(\begin{array}{cc}
			0.275 &0.030\\
			0.030  & 0.004
			\end{array}\right)$
			&\multirow{6}{*}{229.193}&\multirow{6}{*}{6.438}&\multirow{6}{*}{3-G}         \\
			&0.508				&(1.337,0.040)
			&$\left(\begin{array}{cc}
			0.411 &-0.013\\
			-0.013  & 0.016
			\end{array}\right)$
			&&&\\
			&0.466				&(2.002,0.149)
			&$\left(\begin{array}{cc}
			0.140 &-0.009\\
			-0.009  & 0.353
			\end{array}\right)$
			&&&\\
			\hline
			\multirow{12}{*}{\bf Rest}& 0.229              &(0.388, -0.074)
			&  $\left(\begin{array}{cc}
			0.492 &-0.139\\
			-0.139  & 0.083
			\end{array}\right)$
			&\multirow{4}{*}{365.236}&\multirow{4}{*}{0}  &\multirow{4}{*}{2-G}        \\
			&0.771				&(1.357, 0.103)
			& $\left(\begin{array}{cc}
			0.283 &-0.005\\
			-0.005  & 0.010
			\end{array}\right)$
			&&& \\
			&&&&&&\\
			&0.454				&(0.745,0.043)
			&$\left(\begin{array}{cc}
			0.566 &-0.056\\
			-0.056  & 0.021
			\end{array}\right)$
			&\multirow{6}{*}{368.645}&\multirow{6}{*}{3.409}&\multirow{6}{*}{3-G}         \\
			&0.055				&(0.866,-0.465)
			&$\left(\begin{array}{cc}
			0.203 &-0.044\\
			-0.044  & 0.034
			\end{array}\right)$
			&&&\\
			&0.491				&(1.526,0.139)
			&$\left(\begin{array}{cc}
			0.165 &-0.007\\
			-0.007  & 0.006
			\end{array}\right)$
			&&&\\
			\hline

		\end{tabular}
	}
	\renewcommand{\arraystretch}{1}
	\label{tab:gmm_tab_plcpl}

\end{table}

\begin{table}[!ht]\centering
	\renewcommand{\arraystretch}{1.5}
	\caption{Sample Table.~\ref{tab:gmm_tab_plcpl}, but all values are calculated by using PL only.}
	\scalebox{0.55}{
		\begin{tabular}{lcccccc}\hline
			&&&\textbf{PL Only}&&&\\ \hline
			Frame &  $\omega$   &$\bm{\mu}$          & $\bm{\Sigma}$   &BIC  &${\Delta}$	   &Model     \\
			\hline
			\multirow{12}{*}{\bf Observer}& 0.335              &(1.021, 0.045)
			&  $\left(\begin{array}{cc}
			0.642 &-0.059\\
			-0.059  & 0.0025
			\end{array}\right)$
			&\multirow{4}{*}{189.019}&\multirow{4}{*}{0} &\multirow{4}{*}{2-G}        \\
			&0.665				&(1.887, 0.127)
			& $\left(\begin{array}{cc}
			0.194 &-0.004\\
			-0.004  & 0.006
			\end{array}\right)$
			&& \\
			&&&&&&\\
			& 0.026           &(-0.538,0.353)
			&$\left(\begin{array}{cc}
			0.273 &0.029\\
			0.029  & 0.004
			\end{array}\right)$
			&\multirow{6}{*}{196.689}&\multirow{6}{*}{7.670}&\multirow{6}{*}{3-G}         \\
			&0.516				&(1.356,0.049)
			&$\left(\begin{array}{cc}
			0.424 &-0.008\\
			-0.008  & 0.424
			\end{array}\right)$
			&&&\\
			&0.458				&(1.993,0.143)
			&$\left(\begin{array}{cc}
			0.138 &-0.007\\
			-0.007  & 0.005
			\end{array}\right)$
			&&&\\
			\hline
			\multirow{12}{*}{\bf Rest}& 0.417              &(0.697, 0.056)
			&  $\left(\begin{array}{cc}
			0.596 &-0.047\\
			-0.047  & 0.022
			\end{array}\right)$
			&\multirow{4}{*}{200.450}&\multirow{4}{*}{0}  &\multirow{4}{*}{2-G}        \\
			&0.583				&(1.448, 0.131)
			& $\left(\begin{array}{cc}
			0.191 &-0.005\\
			-0.005  & 0.006
			\end{array}\right)$
			&&& \\
			&&&&&&\\
			& 0.125           &(0.060, 0.074)
			&$\left(\begin{array}{cc}
			0.567 &-0.056\\
			-0.056  & 0.021
			\end{array}\right)$
			&\multirow{6}{*}{222.609}&\multirow{6}{*}{22.159}&\multirow{6}{*}{3-G}         \\
			&0.479				&(1.104,0.061)
			&$\left(\begin{array}{cc}
			0.388 &-0.017\\
			-0.017  & 0.020
			\end{array}\right)$
			&&&\\
			&0.396				&(1.511,0.156)
			&$\left(\begin{array}{cc}
			0.156 &-0.008\\
			-0.008  & 0.004
			\end{array}\right)$
			&&&\\
			\hline

		\end{tabular}
	}
	\renewcommand{\arraystretch}{1}
	\label{tab:gmm_tab_pl}

\end{table}

\begin{figure*}[!ht]
	\centering
	\includegraphics[width=1.0\columnwidth]{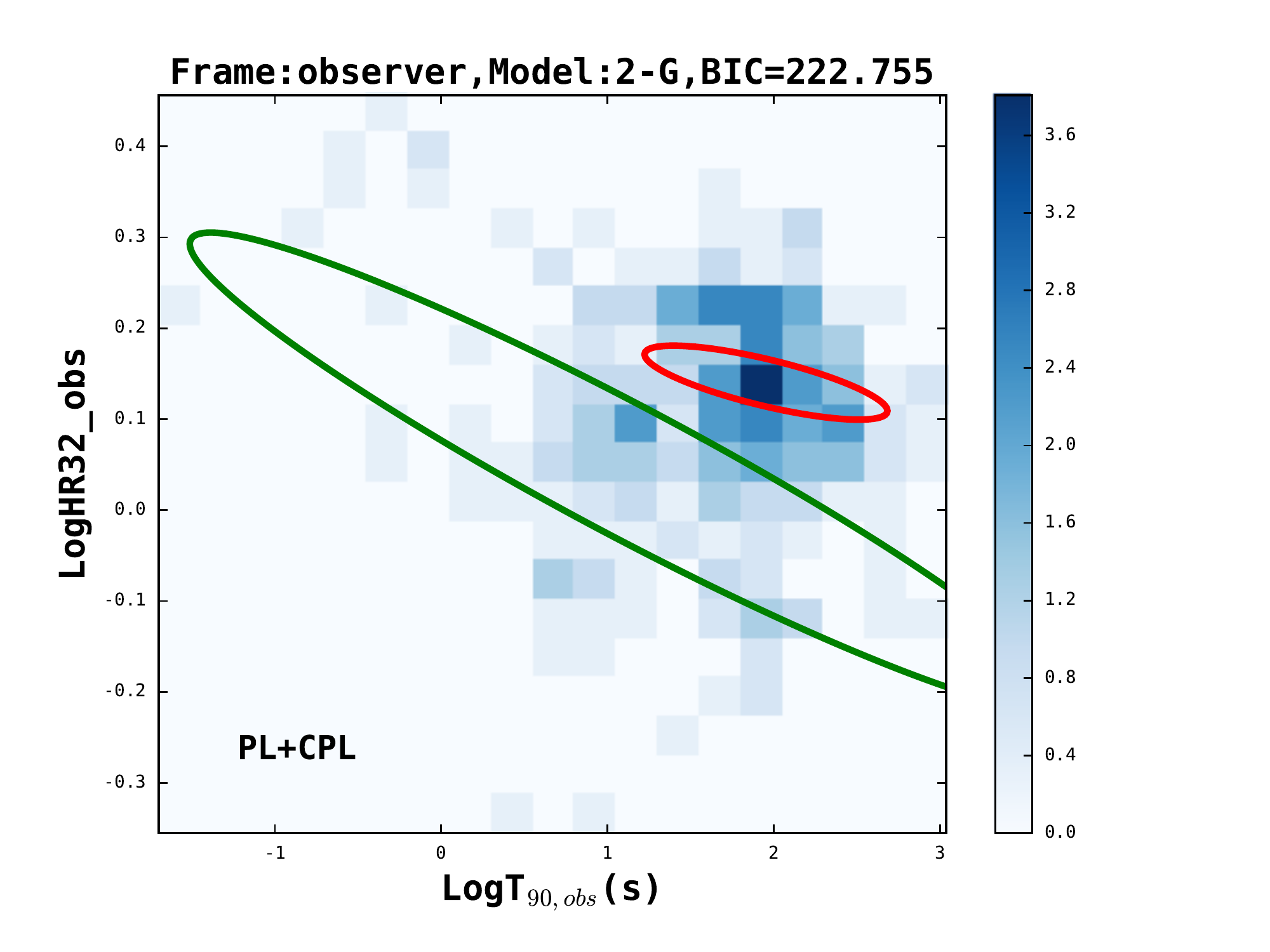}\includegraphics[width=1.0\columnwidth]{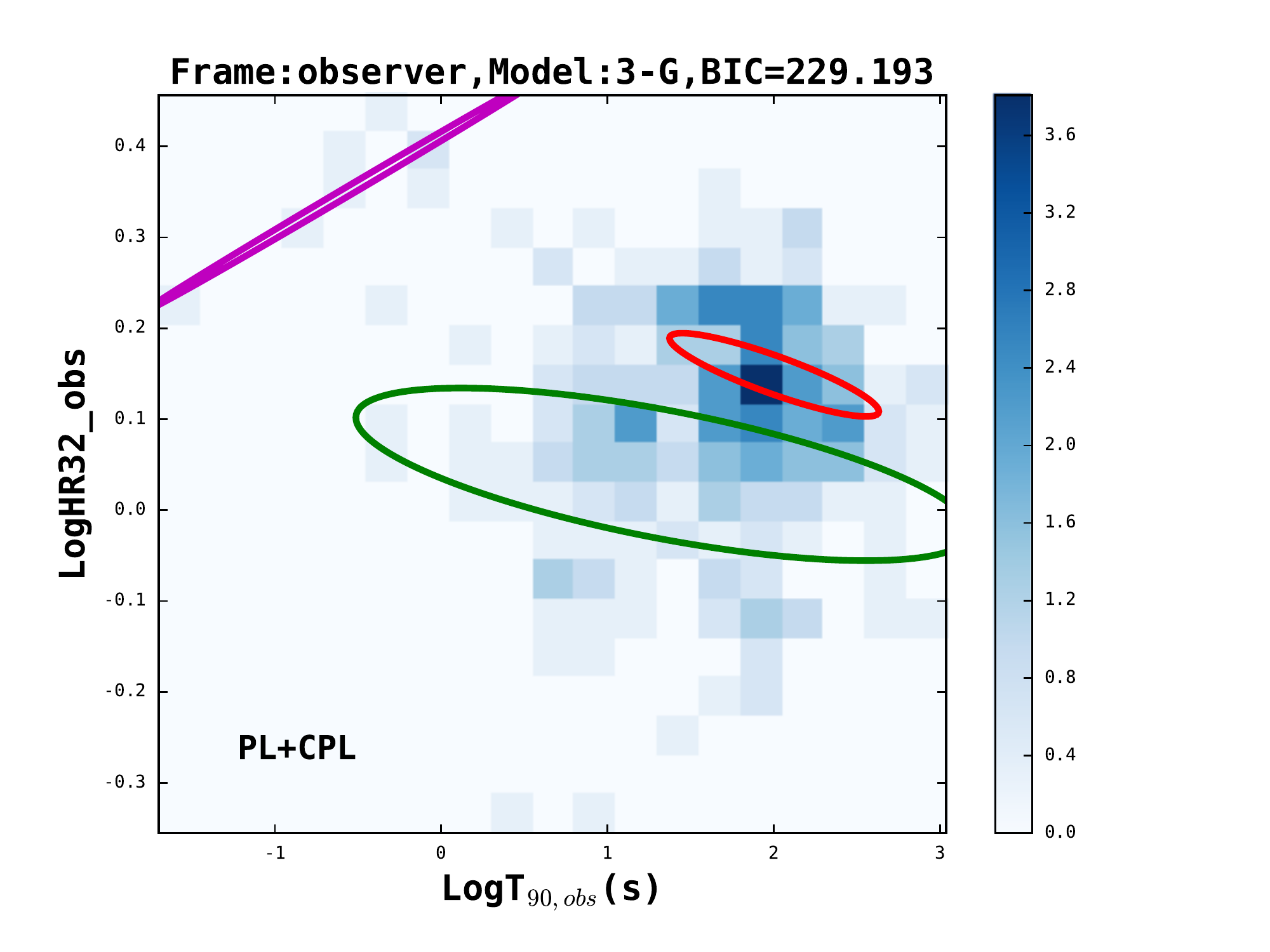}
	\includegraphics[width=1.0\columnwidth]{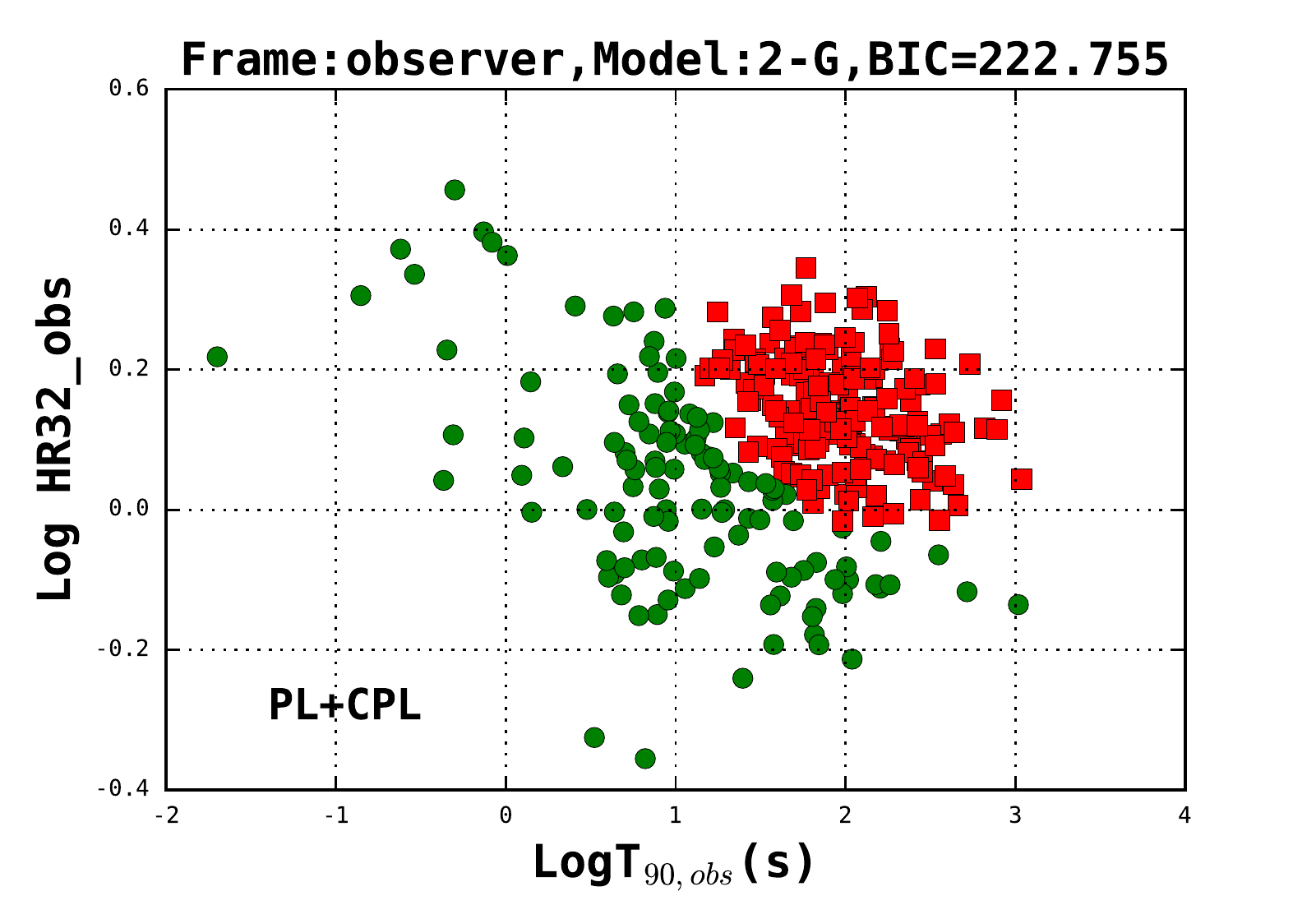}\includegraphics[width=1.0\columnwidth]{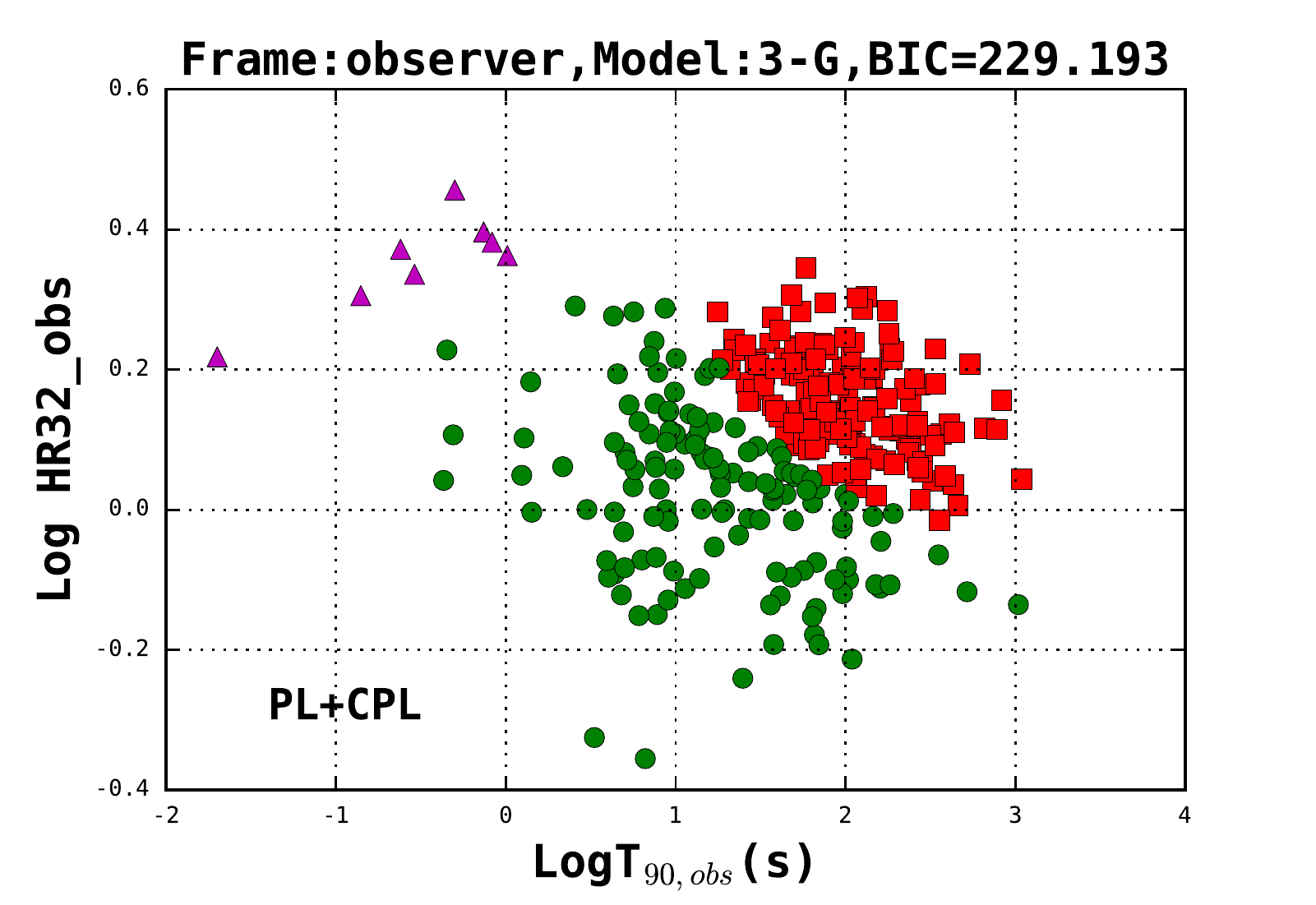}
	\caption{GMM analysis results in the observer frame, calulated by
	mixed spectrum model, PL or CPL. Two figures in the upper panel
	are the 1$\sigma$ ellipses of the best GMM 2-G model (upper-left) and 3-G
	model (upper-right), respectively. The background in these two figures are
	2 dimensional histogram following Knuth bin rule \citep{knuth}. The lower
	two figures are the distributions of our datasets. The different symbols
	with different colors represent those GRBs corresponding  to different
	Gaussian components.}\label{fig_obs_plcpl}
\end{figure*}

\begin{figure*}[!ht]
	\centering
	\includegraphics[width=1.0\columnwidth]{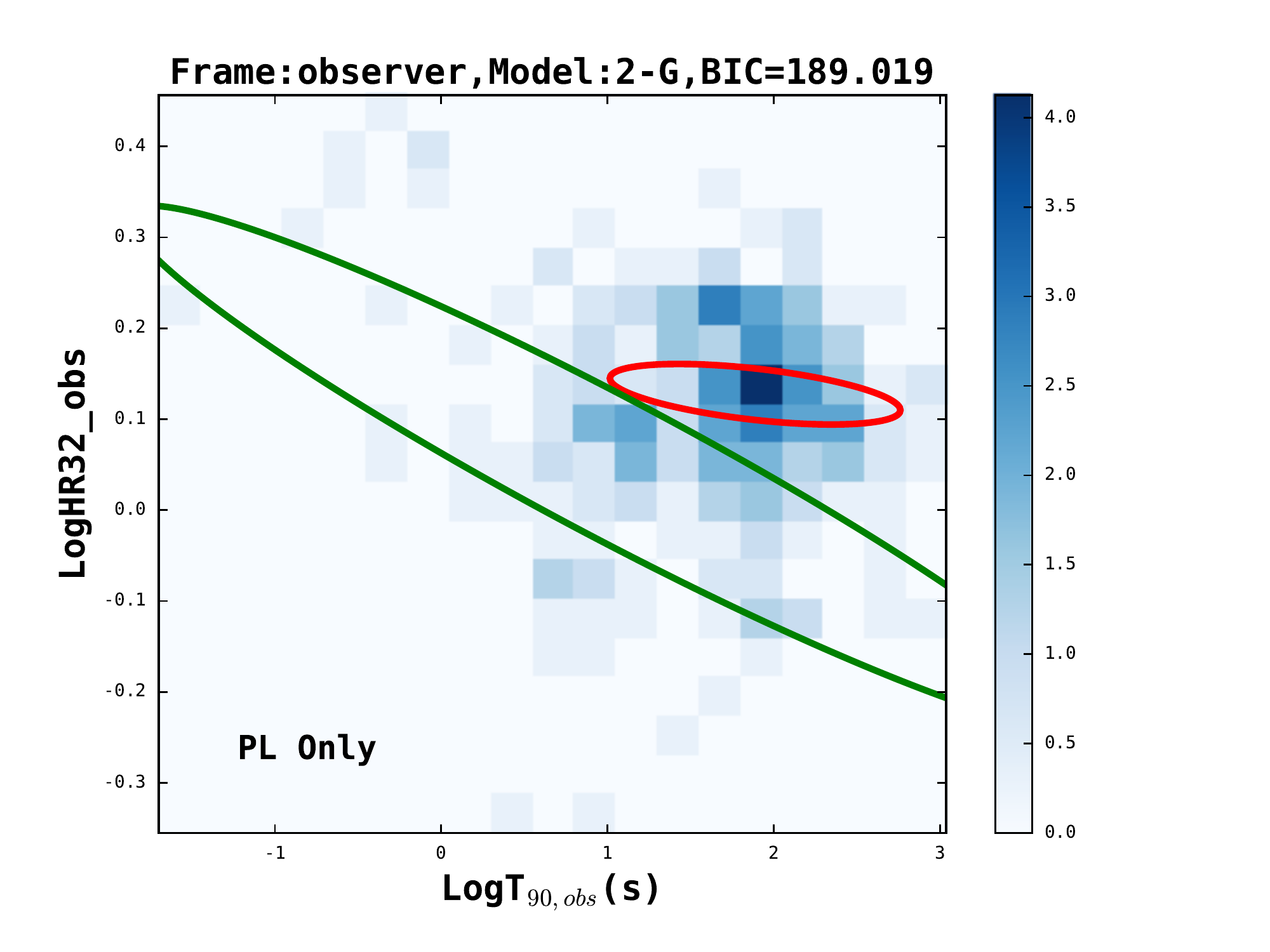}\includegraphics[width=1.0\columnwidth]{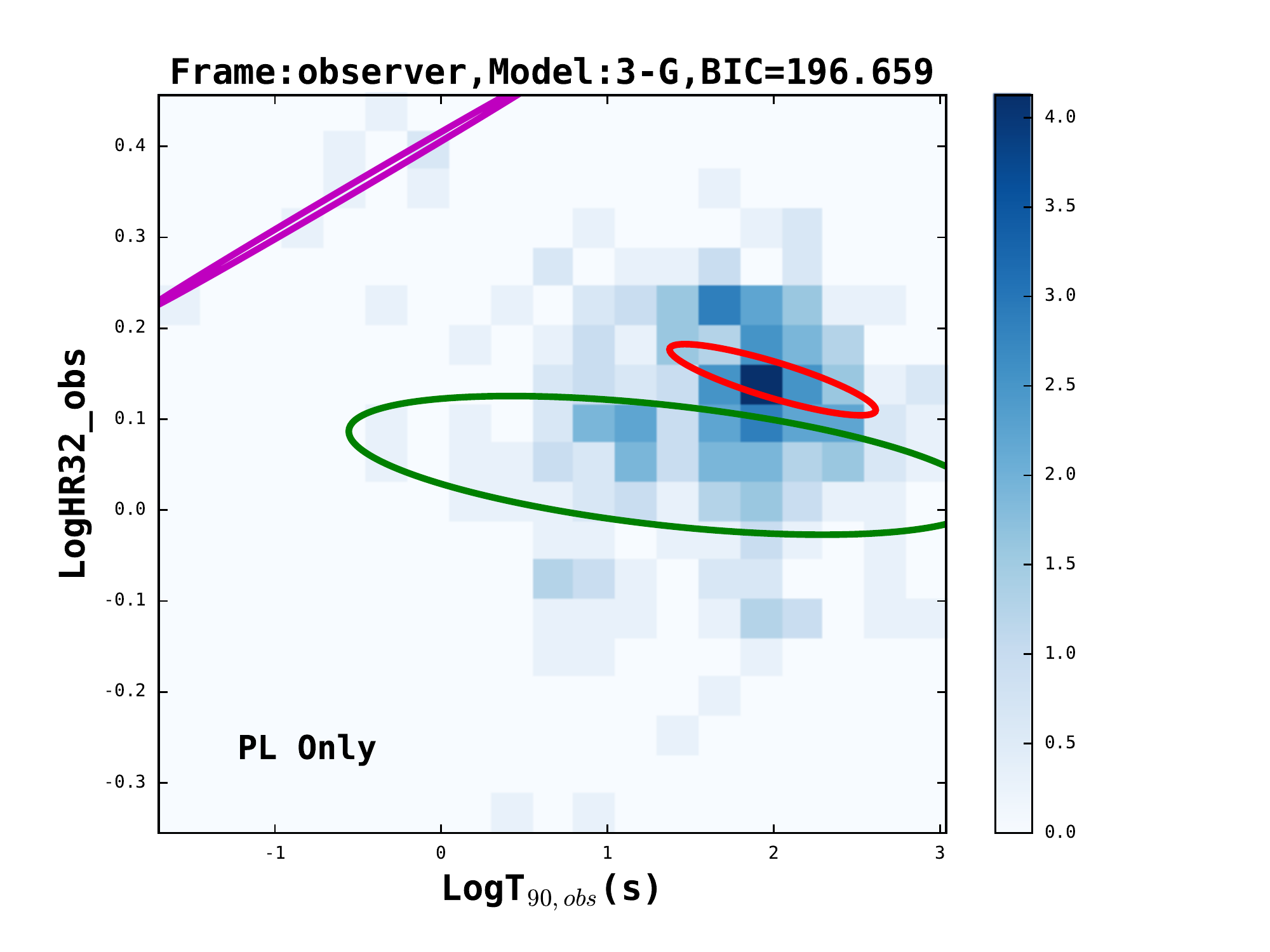}
	\includegraphics[width=1.0\columnwidth]{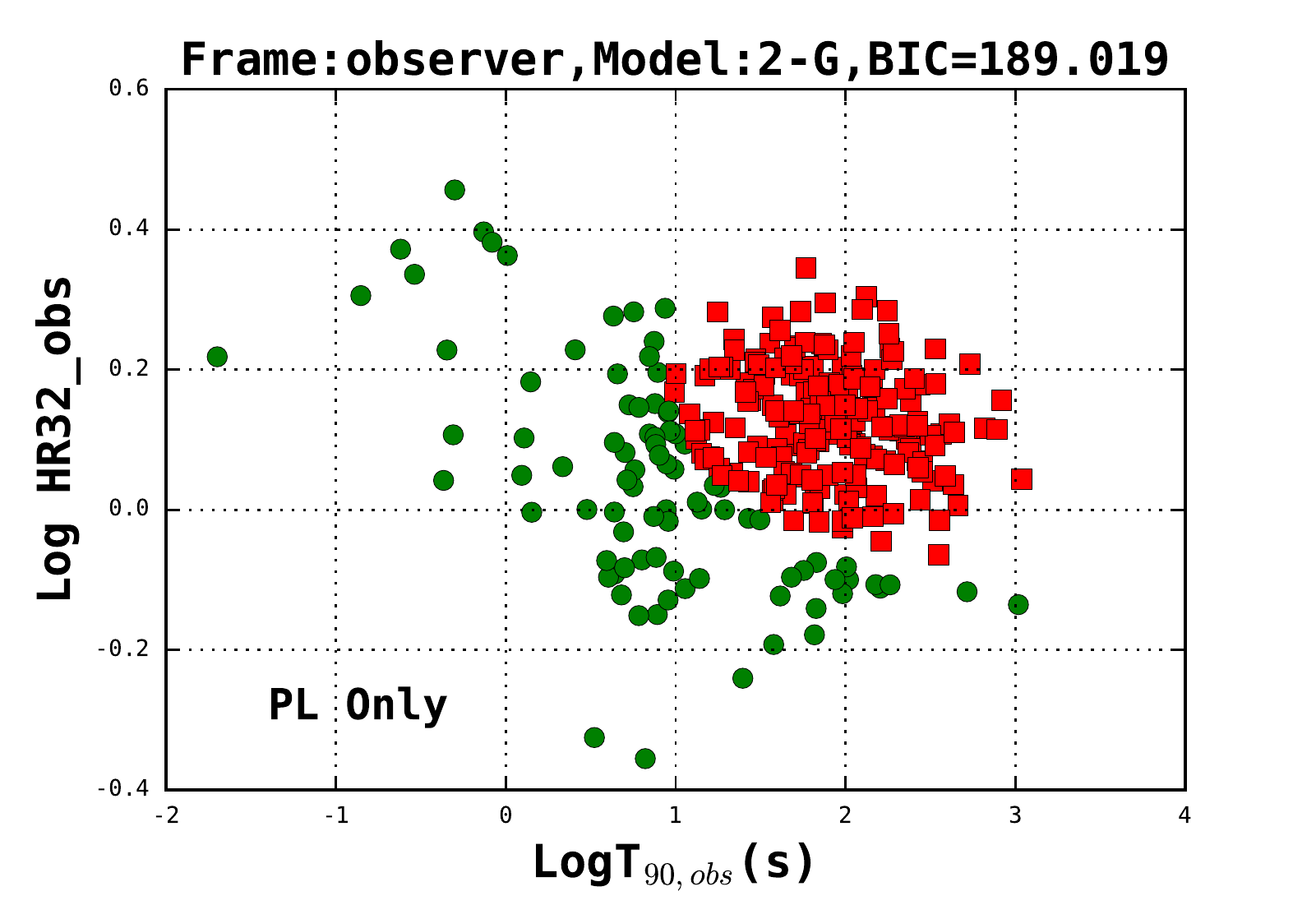}\includegraphics[width=1.0\columnwidth]{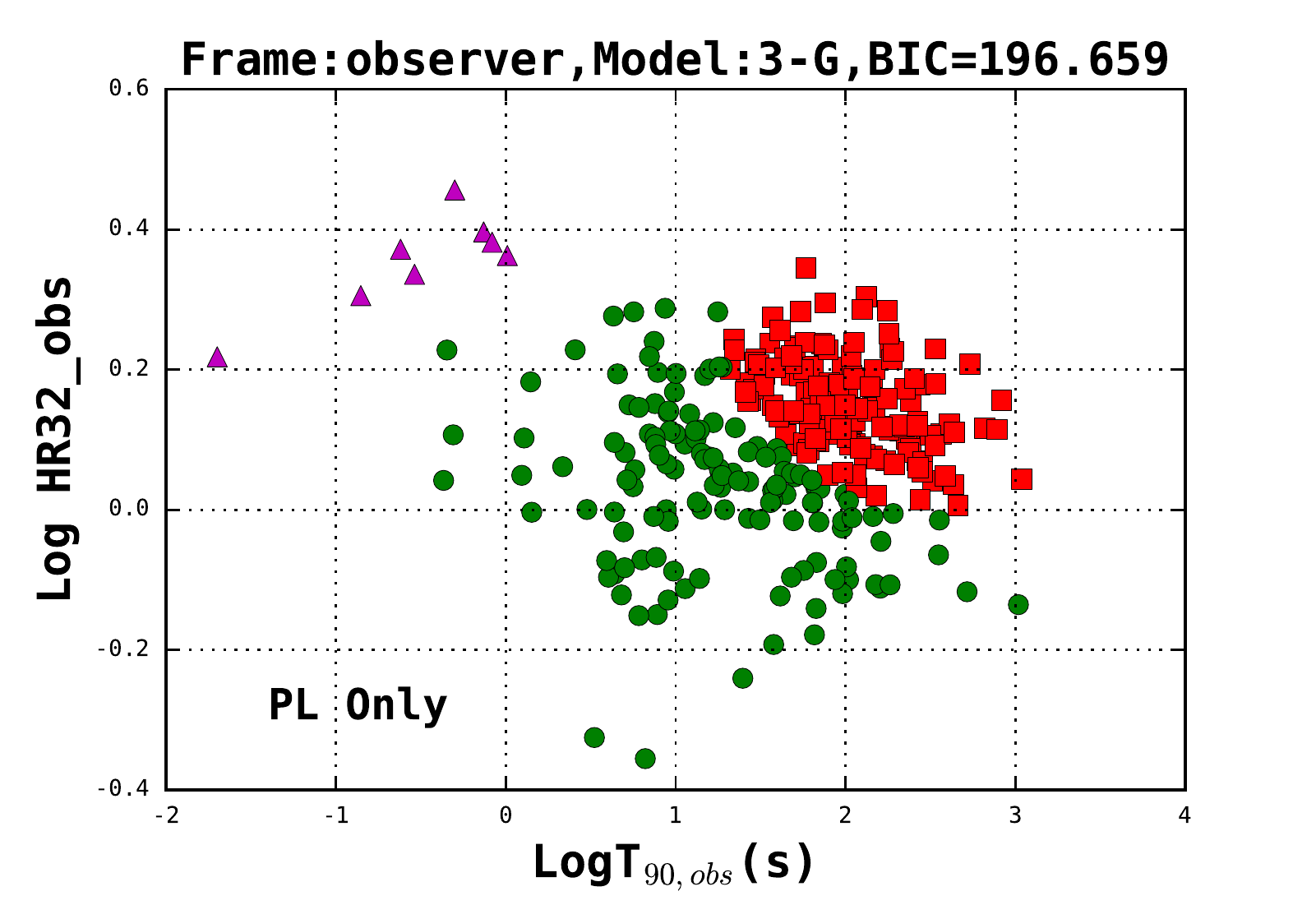}
	\caption{Same as Fig.~\ref{fig_obs_plcpl}, but calculated by using
	PL only.}\label{fig_obs_pl}
\end{figure*}


\begin{figure*}[!ht]
	\centering
	\includegraphics[width=1.0\columnwidth]{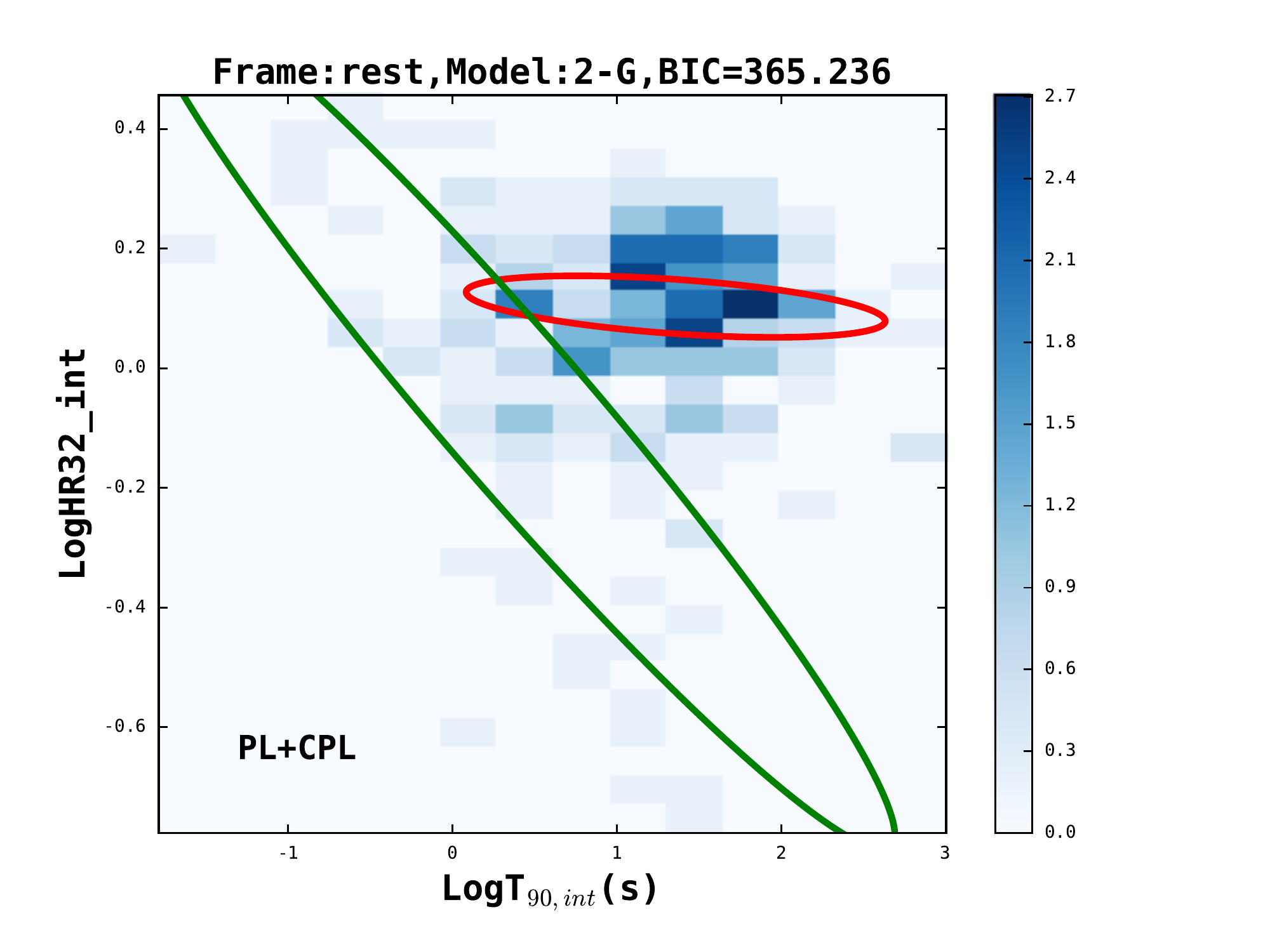}\includegraphics[width=1.0\columnwidth]{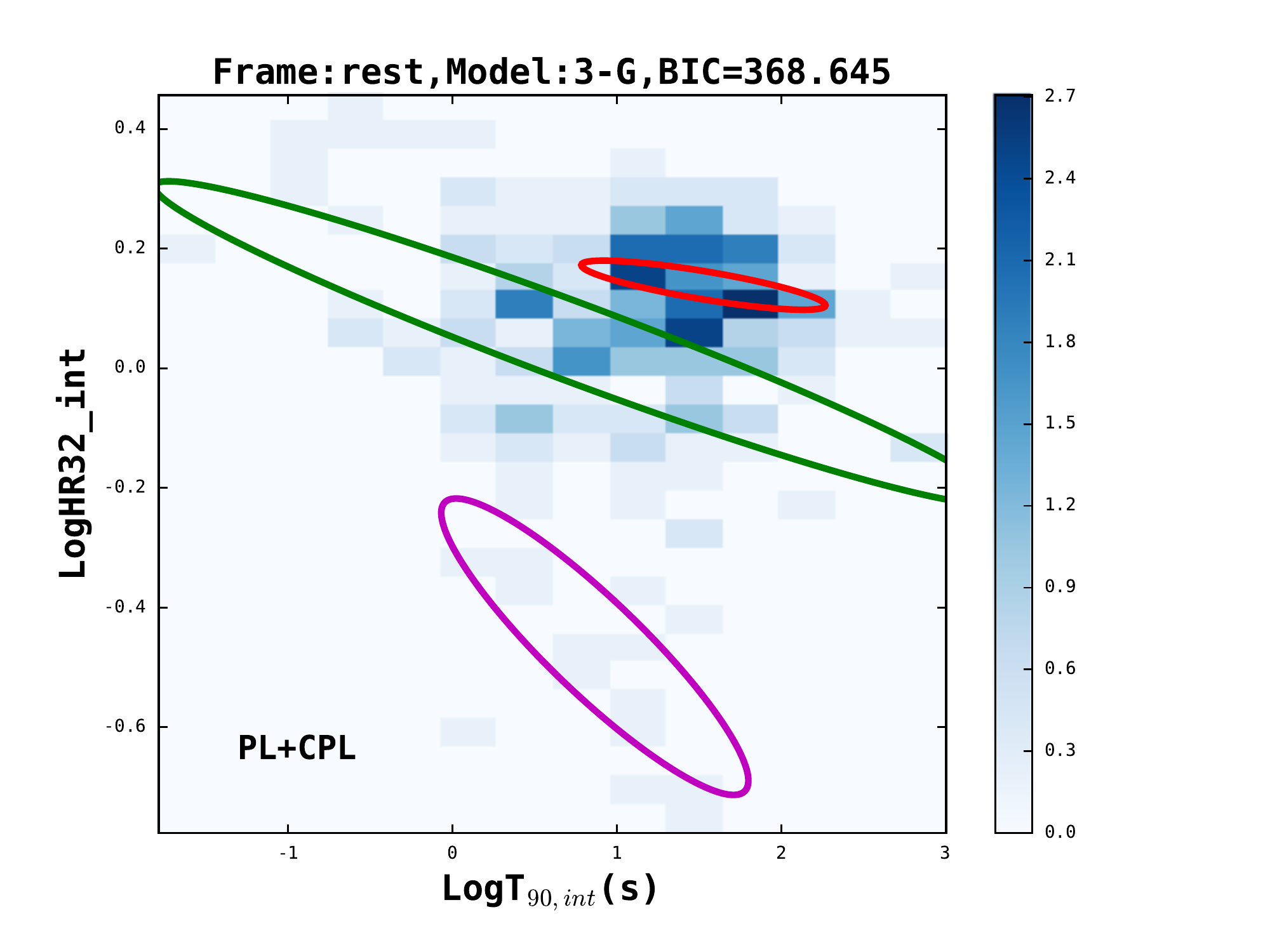}\\
	\includegraphics[width=1.0\columnwidth]{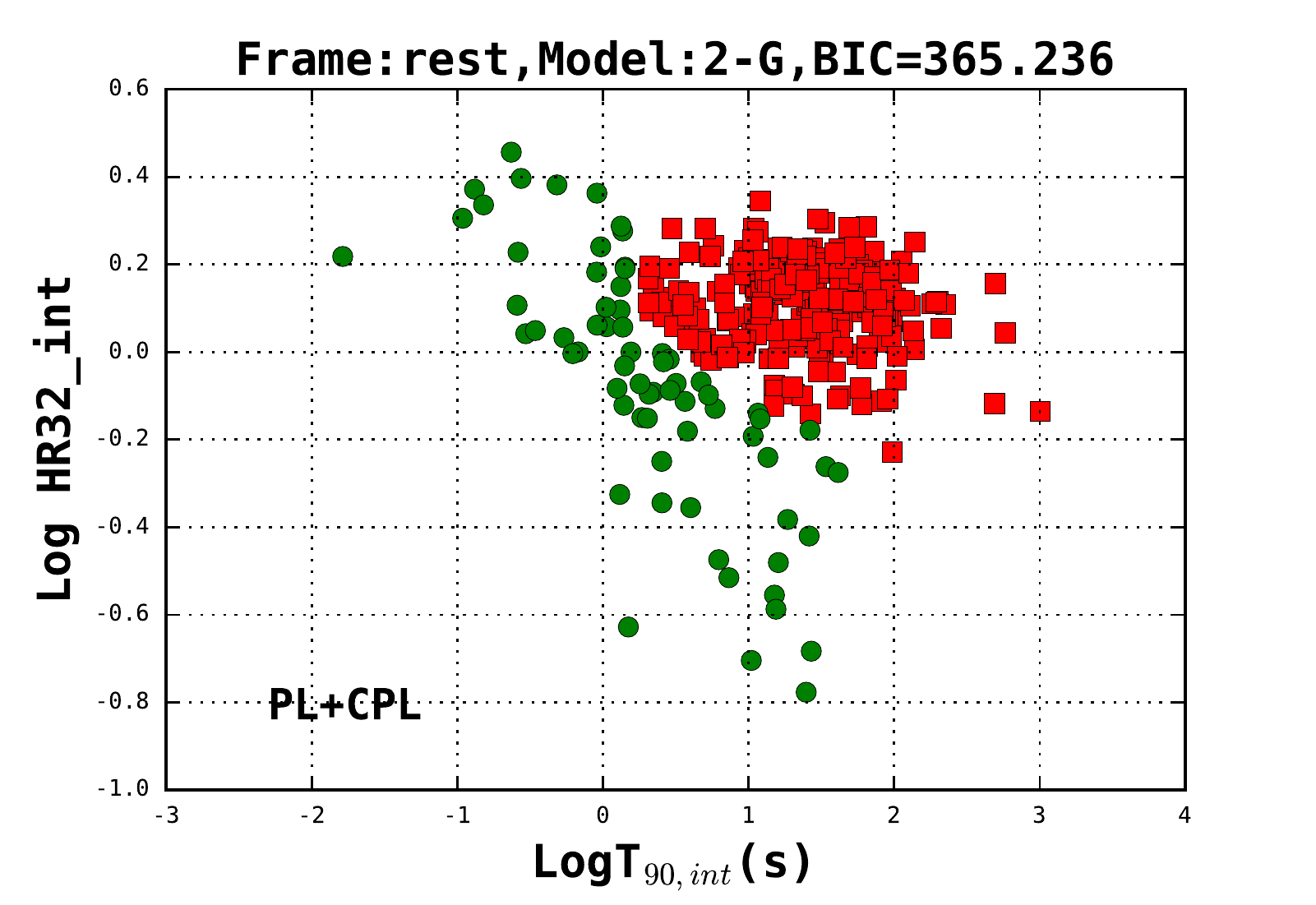}\includegraphics[width=1.0\columnwidth]{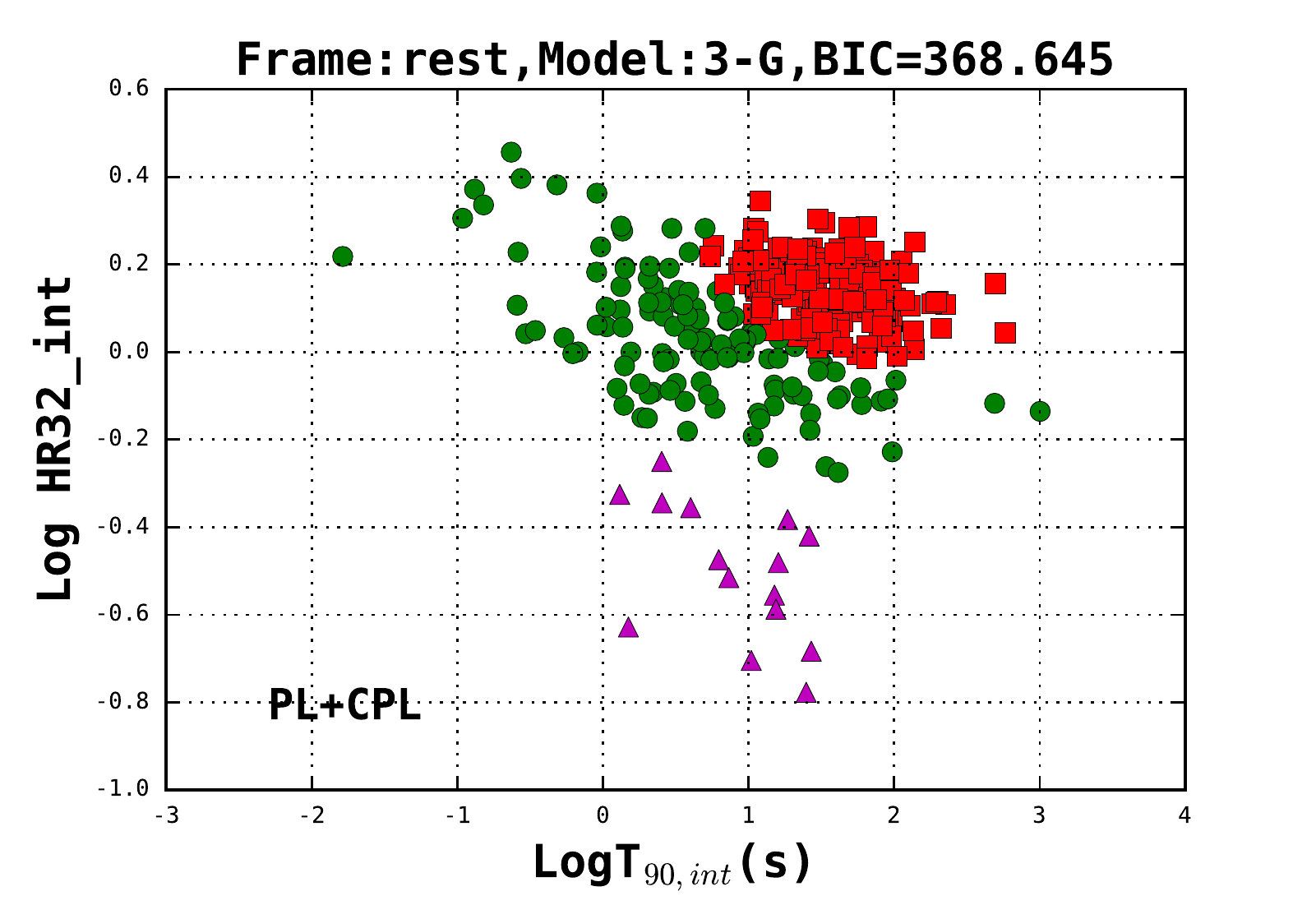}
	\caption{GMM analysis results in the rest frame, calculated by
	using mixed spectrum model, PL or CPL. Order of models are the
	same as in Fig.~\ref{fig_obs_plcpl}}\label{fig_rest_plcpl}
\end{figure*}

\begin{figure*}[!ht]
	\centering
	\includegraphics[width=1.0\columnwidth]{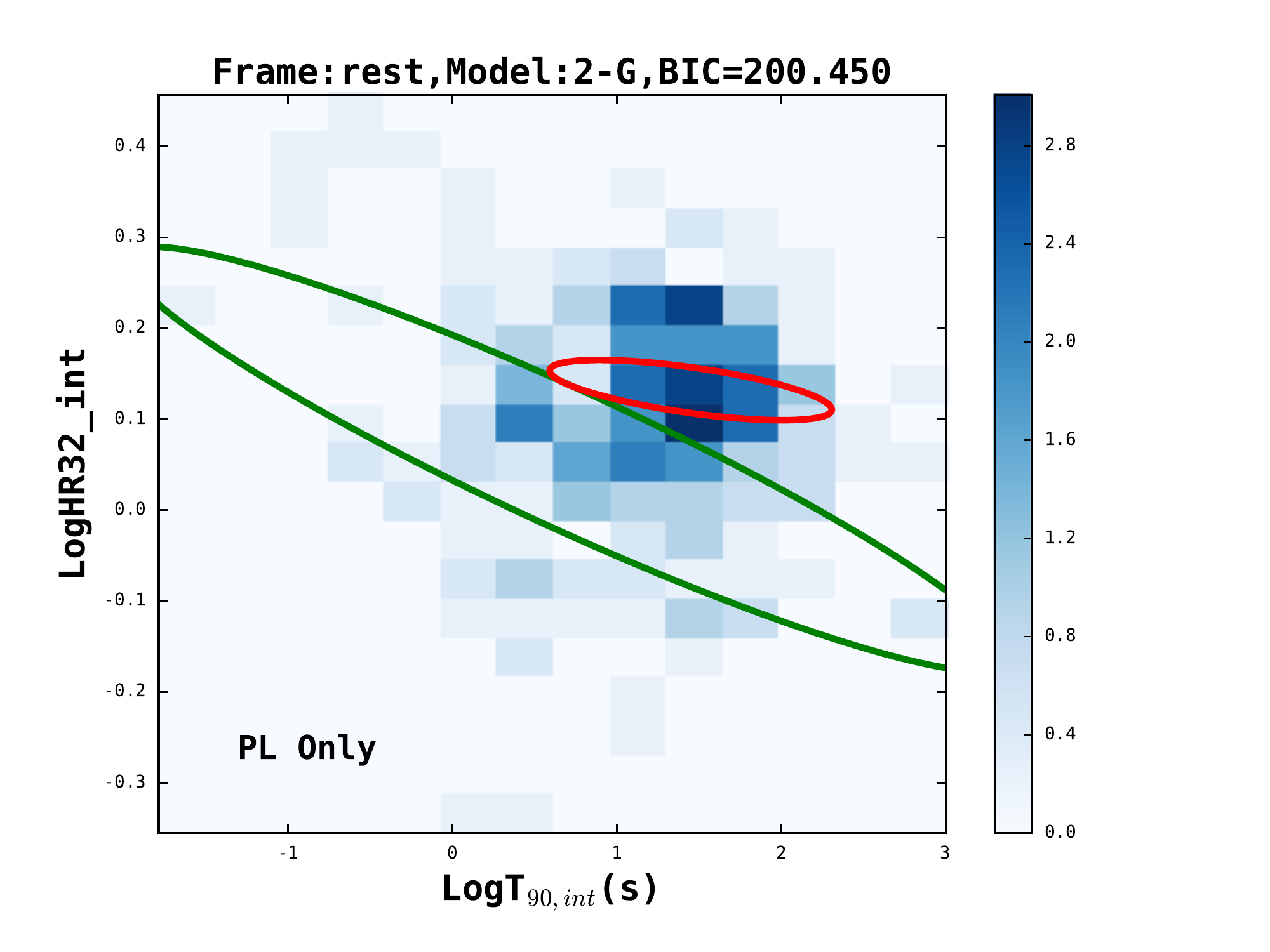}\includegraphics[width=1.0\columnwidth]{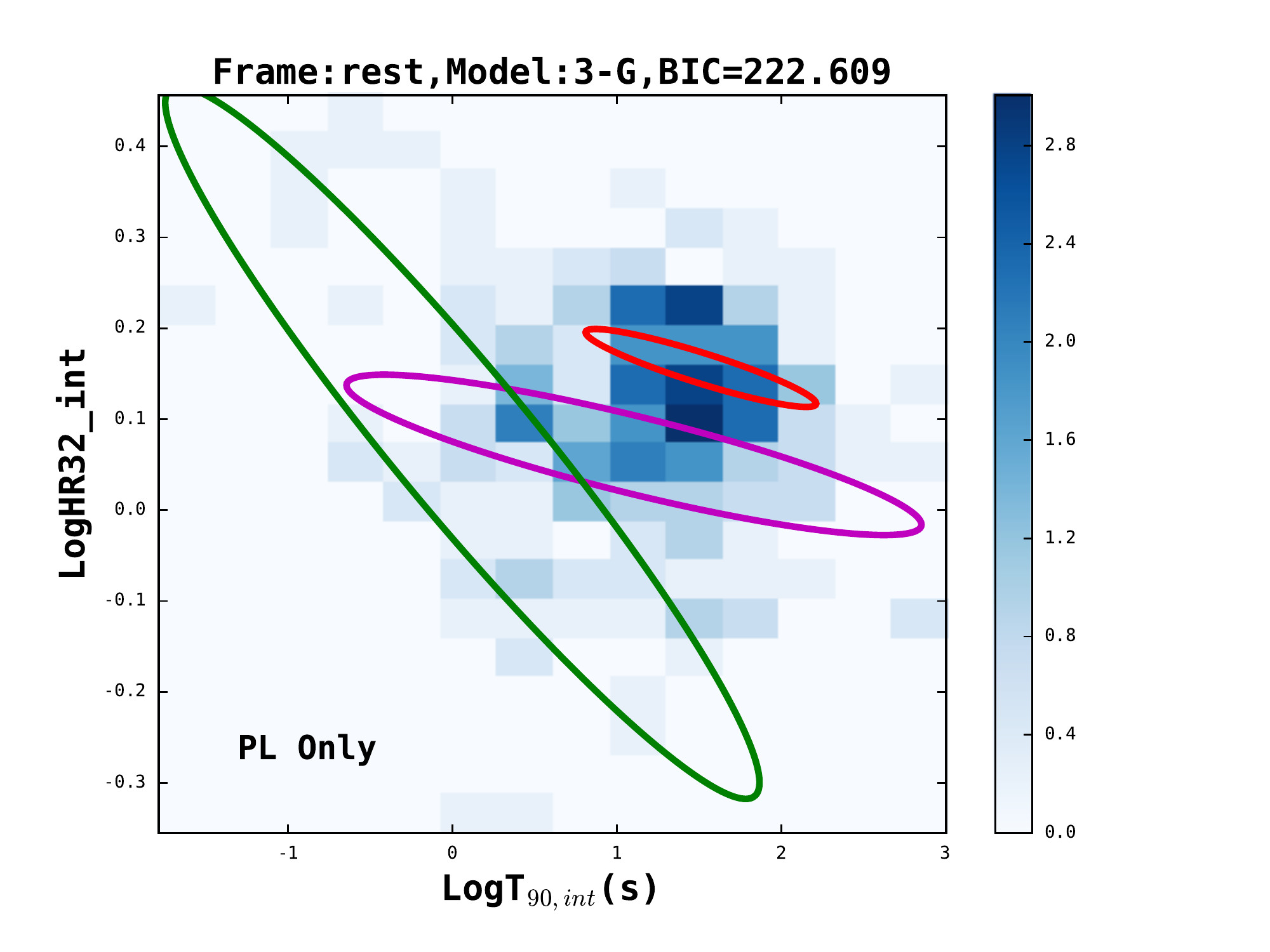}\\
	\includegraphics[width=1.0\columnwidth]{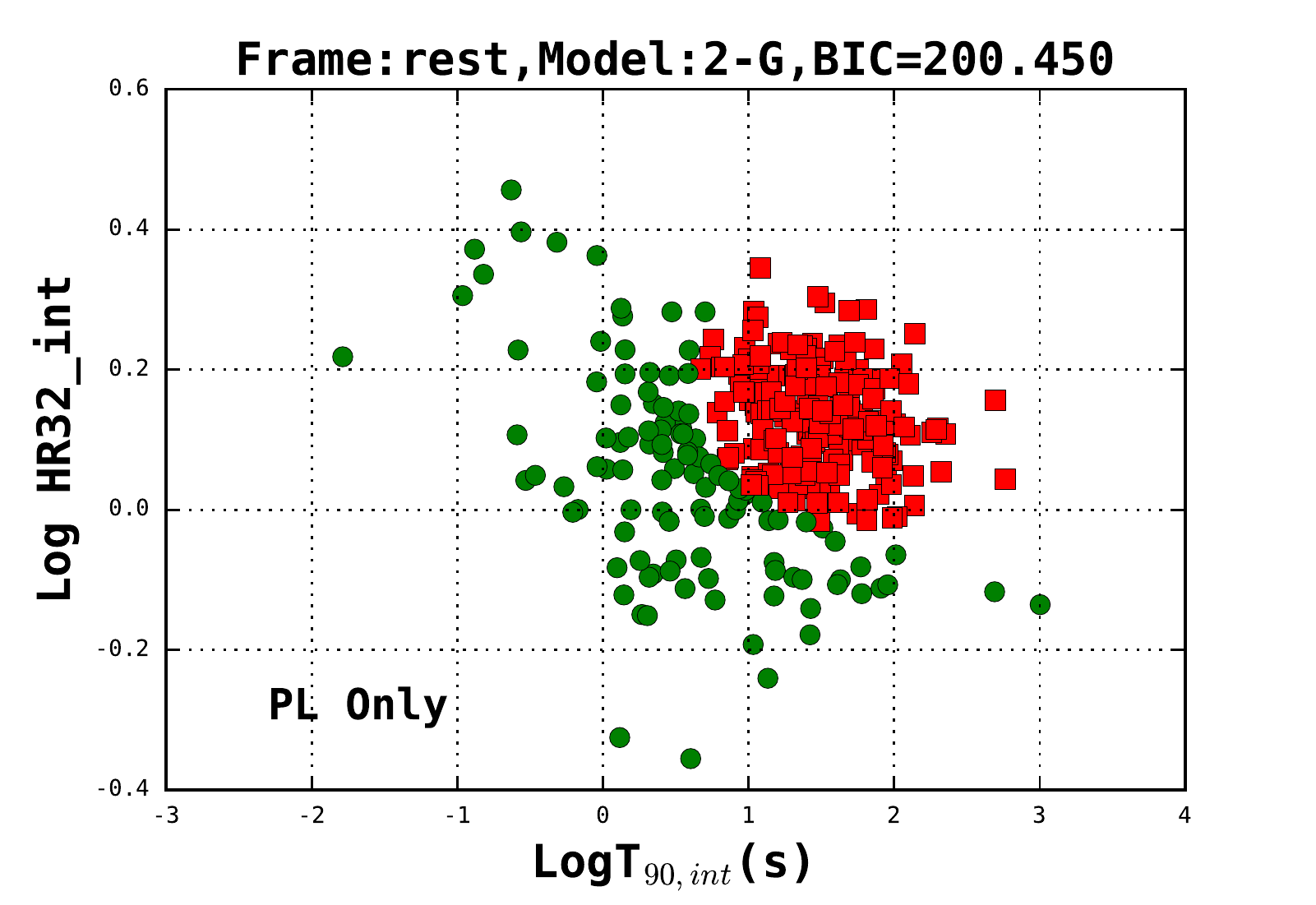}\includegraphics[width=1.0\columnwidth]{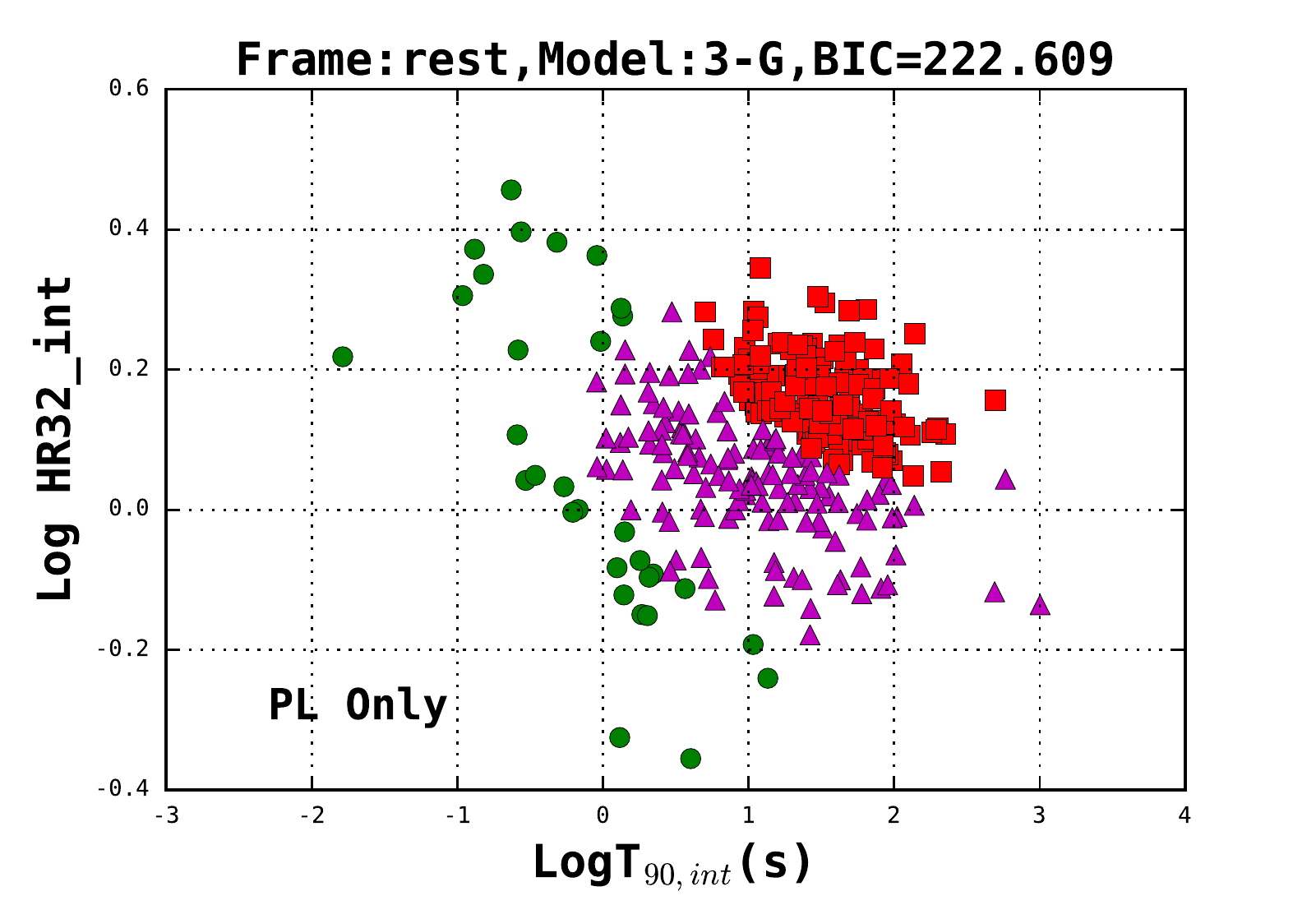}
	\caption{Same as Fig.~\ref{fig_rest_plcpl}, but calculated by using PL
	only}\label{fig_rest_pl}
\end{figure*}


\begin{figure}[!ht]
	\centering
	\includegraphics[width=1\columnwidth]{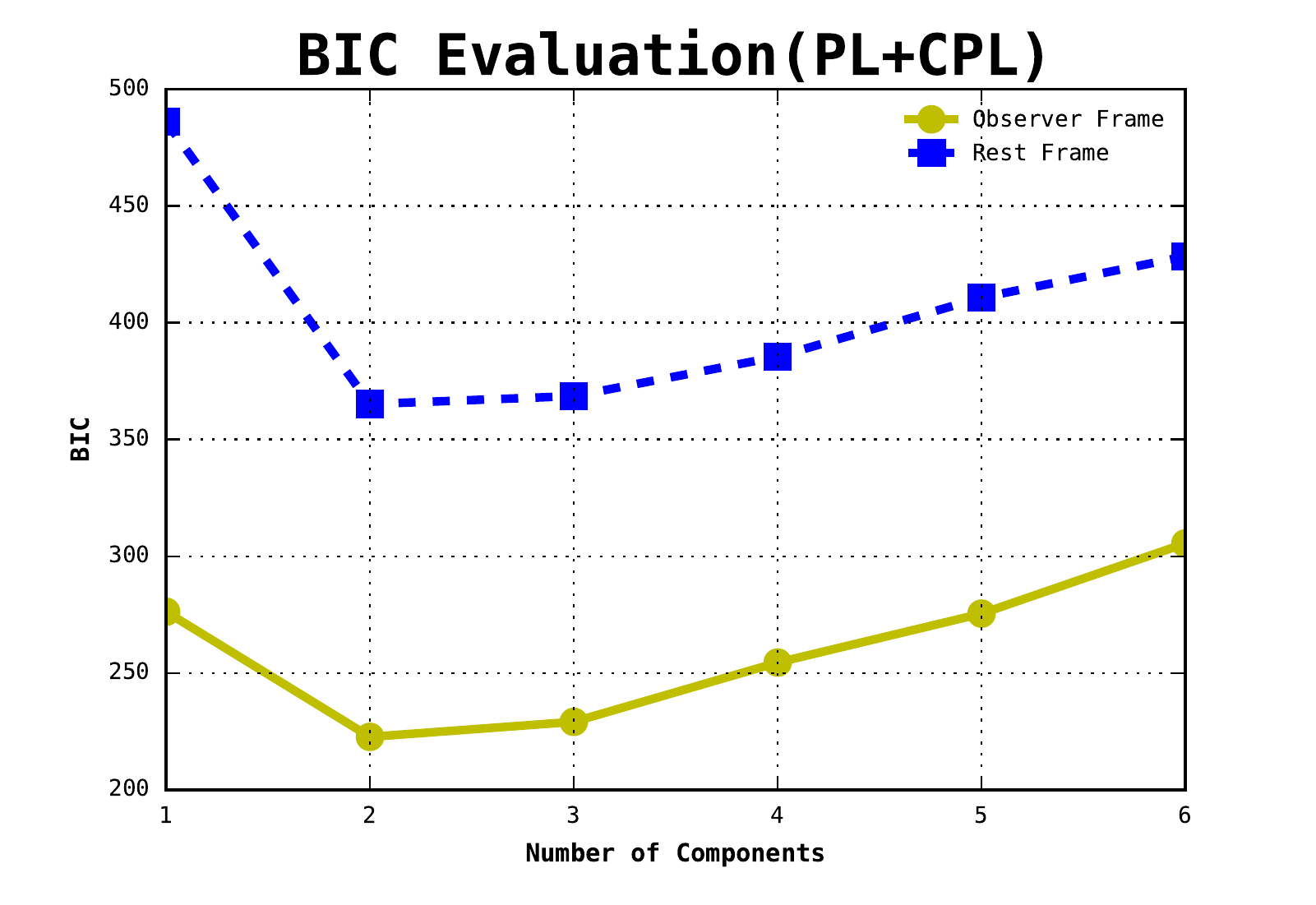}
	\caption{BIC evaluation for models with different number of
	Gaussian components under the calculation by using mixed spectrum model.
	Yellow solid line is for the observer frame and blue dashed
	line is for the rest frame.} \label{bic_plcpl}
\end{figure}
\begin{figure}[!ht]
	\centering
	\includegraphics[width=1\columnwidth]{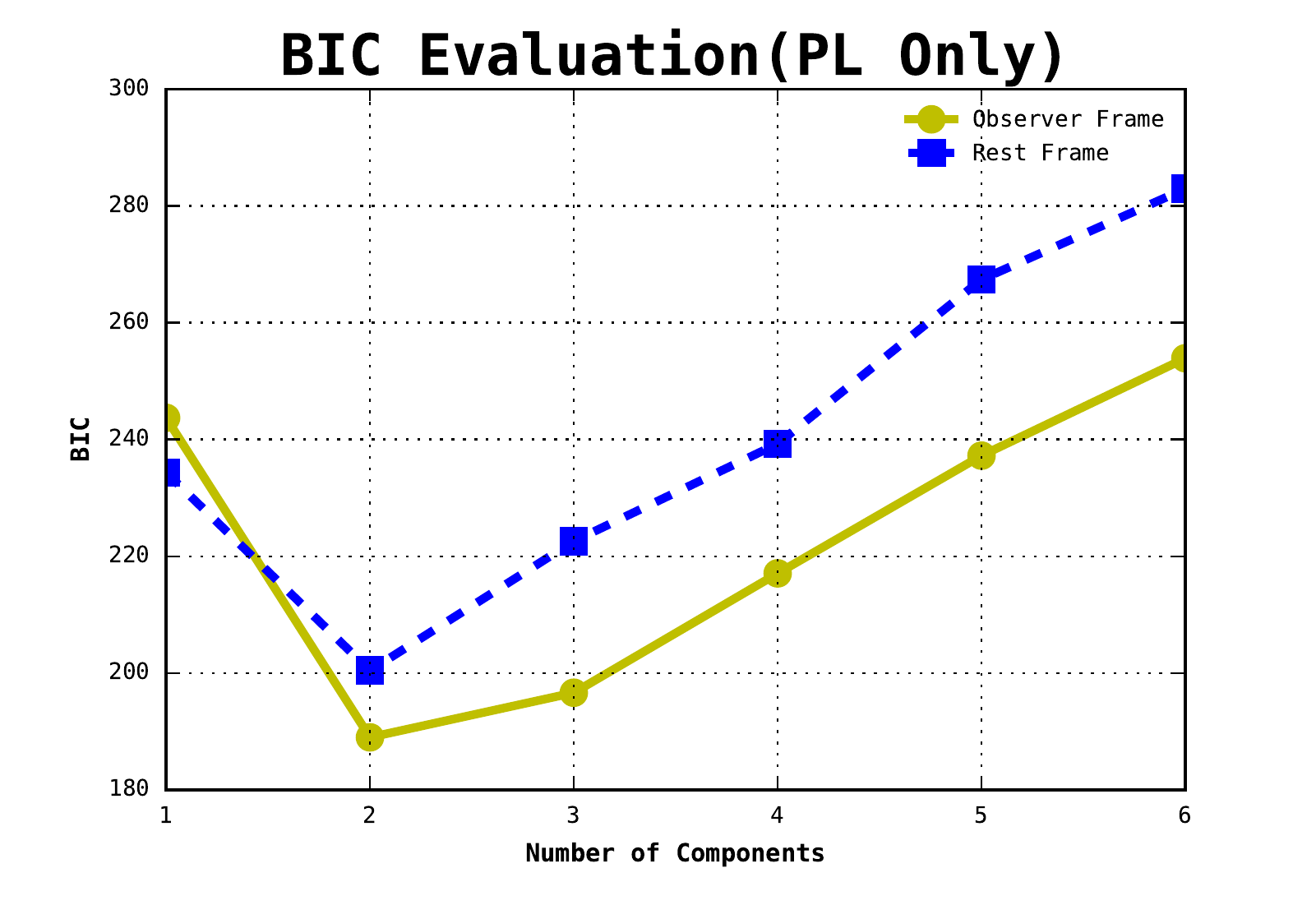}
	\caption{Same as Fig.~\ref{bic_plcpl}, but under the calculation by
	using the PL only.} \label{bic_pl}
\end{figure}

\section{Results}\label{sect3}

Using the GMM method, we estimated the $\log{T_{90}}$ vs $\log{HR_{32}}$ density
distribution of 296 redshift-known \textit{Swift} GRBs in both observer and
rest frames.  Our results are displayed in Fig.~\ref{fig_obs_plcpl} and
\ref{fig_rest_plcpl} under the mixed spectrum model(marked as ``PL+CPL''
in the figures), Fig.~\ref{fig_obs_pl} and \ref{fig_rest_pl} under PL
only(marked as ``PL Only'' in the figures). The optimized parameters of
different models are listed in Table~\ref{tab:gmm_tab_plcpl} and
Table~\ref{tab:gmm_tab_pl}, respectively.


In the observer frame, as shown in Fig.~\ref{fig_obs_plcpl}, the
minimum BIC value obtained with the mixed spectrum model for 2 Gaussian
components(2-G) model is $\sim$223, about 6.5
smaller than $\sim$229 for 3 Gaussian components(3-G) model(see
Table~\ref{tab:gmm_tab_plcpl}). This indicates that 2 components are better
enough to describe the two dimensional distribution of
$T_{90,obs}\sim{HR}_{32,obs}$ in the observer frame. Similar results can also be
obtained under the calculation by using PL only in Fig.~\ref{fig_obs_pl} and
Table~\ref{tab:gmm_tab_pl}, where minimum BIC value for
2-G is $\sim$ 189, about 7.7 smaller than $\sim$197 for 3-G.

In the rest frame, the intrinsic duration $T_{90,int}$ can be calculated by
$T_{90,int}=\frac{T_{90,obs}}{(1+z)}$.
The intrinsic $HR_{32,int}$ is calculated with the Eq.~\ref{hr_res}.
Once again, we find that the BIC prefers the 2-G model instead of the
3-G model. As shown in
Fig.~\ref{fig_rest_plcpl}, and Table~\ref{tab:gmm_tab_pl}, by mixture spectral
model, the minimum BIC value for 2-G is about 3.4 smaller
than 3-G's BIC. For calculation using PL only, the minimum BIC value 3-G is
about 22 larger than 2-G's BIC. 
Simultaneously, $T_{90,int}$ of the two components move to the smaller 
value, systematically, as listed in Table~\ref{tab:gmm_tab_plcpl} and 
\ref{tab:gmm_tab_pl}.

Fig.~\ref{bic_plcpl} and \ref{bic_pl} show our BIC evaluation of 
different models in both observer and rest frames under two calculation 
condition. One can find that two instead of three or more Gaussian components 
are favoured in both the observer and rest frames.

Our work shows that 2-G model is better than 3-G model in the observer and rest frames, which is inconsistent with some previous results (e.g. \citealt{horvath2010}). For the 3-G model in the observer frame, our analysis returns relative ratios 0.026, 0.508 and 0.466 (PL+CPL) or 0.026, 0.516 and 0.458 (PL only), which are significantly different from 0.079, 0.296 and 0.626 gotten by \cite{horvath2010} (see also \citealt{koen}). Considering the datasets with redshift-known \textit{Swift} GRBs only, the above difference is due to the small fraction of short GRBs in our sample, which makes short GRBs with redshift not statistically significant as a single group possibly. This will be confirmed by a larger population of short GRBs with measured redshift in the future.

\cite{veres} also studied the \textit{Swift}/BAT GRBs and found they favour 
3-component model. They argue that the third component is closely related 
to X-ray Flashes(XRFs). For the mixed spectrum model in 
Fig.~\ref{bic_plcpl}, the 3-G model under the calculation by 
using mixed spectrum model is weakly supported in the rest 
frame and the third component indeed have the softest spectra (see 
Fig.~\ref{fig_rest_plcpl}). This may indicate that the connection between the 
third component and the XRFs might be true.Based on the above analysis in 
the two dimensional case, one can 
conclude that the \textit{Swift} bursts with measured redshift are still better 
to  be classified into two groups, which is in good agreement with some  
previous works (e.g. \citealt{zhangzb2008,qin2013,tarno2015aa,yangeb})

The results of this work can be summarized as.
\begin{enumerate}
	\item \textit{Swift} GRBs are better to describe with two groups other than
	three in either the observer or the rest frame.
	\item This classification criterion is not affected by the spectral form
	too much.
	\item The third class weakly supported in the rest frame may correspond to
	the softest spectra of the XRFs.
	
	\item More reshift-known GRBs are needed to do this kind 
		of research.
\end{enumerate}

\acknowledgments
We thank the anonymous referee for valuable comments and suggestions that led
to an overall improvement of this work.
This work is partly supported by the National Natural Science
Foundation of China (Grant No. U1431126; 11263002; 11311140248; 11203026)
and Provincial Natural Science Foundations (201519; 20134021; 20117006; OP201511).

\clearpage{}

\end{document}